\documentclass[dvipsnames,aps,prl,nobibnotes,twocolumn,superscriptaddress,nolongbibliography]{revtex4-1}

\usepackage[T1]{fontenc}
\usepackage[english]{babel}
\usepackage[utf8]{inputenc}

\usepackage[unicode=true,pdfusetitle,bookmarks=true,bookmarksnumbered=false,bookmarksopen=false,
 breaklinks=false,pdfborder={0 0 0},pdfborderstyle={},backref=false,colorlinks=true]
 {hyperref}
 \hypersetup{
 citecolor=blue,
 urlcolor=blue,
 linkcolor=blue}

\usepackage{array}
\usepackage{units}
\usepackage{amsmath}
\usepackage{cancel}
\usepackage{mathrsfs}
\usepackage{amssymb}
\usepackage{graphicx}
\usepackage{stackrel}
\usepackage{babel}
\usepackage{xparse}
\usepackage{soul}
\usepackage[final]{pdfpages}
\usepackage{comment}
\usepackage{tcolorbox}

\newcommand{\new}[1]{{\color{black} {#1}}}

\makeatletter
\AtBeginDocument{\let\LS@rot\@undefined}
\makeatother

\begin{document}
\title{Networks with many structural scales: a Renormalization Group perspective}
\author{Anna Poggialini}
\affiliation{Dipartimento di Fisica Universit\`a ``Sapienza”, P.le
  A. Moro, 2, I-00185 Rome, Italy.}
\affiliation{`Enrico Fermi' Research Center (CREF), Via Panisperna 89A, 00184 - Rome, Italy}
\author{Pablo Villegas}
\email{pablo.villegas@cref.it}
\affiliation{`Enrico Fermi' Research Center (CREF), Via Panisperna 89A, 00184 - Rome, Italy}
\affiliation{Instituto Carlos I de F\'isica Te\'orica y Computacional, Univ. de Granada, E-18071, Granada, Spain.}
\author{Miguel A. Mu\~noz}
\affiliation{Departamento de Electromagnetismo y F\'isica de la Materia, Universidad de Granada, Granada 18071, Spain}
\affiliation{Instituto Carlos I de F\'isica Te\'orica y Computacional, Univ. de Granada, E-18071, Granada, Spain.}
\author{Andrea Gabrielli}
\affiliation{`Enrico Fermi' Research Center (CREF), Via Panisperna 89A, 00184 - Rome, Italy}
\affiliation{Dipartimento di Ingegneria Civile, Informatica e delle Tecnologie Aeronautiche, Universit\`a degli Studi “Roma Tre”, Via Vito Volterra 62, 00146 - Rome, Italy.}
\affiliation{Istituto dei Sistemi Complessi (ISC) - CNR, Rome, Italy.}
          
\begin{abstract}
Scale invariance profoundly influences the dynamics and structure of complex systems, spanning from critical phenomena to network architecture. Here, we propose a precise definition of scale-invariant networks by leveraging the concept of a constant entropy-loss rate across scales in a renormalization-group coarse-graining setting. This framework enables us to differentiate between scale-free and scale-invariant networks, revealing distinct characteristics within each class. Furthermore, we offer a comprehensive inventory of genuinely scale-invariant networks, both natural and artificially constructed, demonstrating, e.g., that the human connectome exhibits notable features of scale invariance. Our findings open new avenues for exploring the scale-invariant structural properties crucial in biological and socio-technological systems.
\end{abstract}

\maketitle   
The network paradigm captures essential attributes of real-world complex systems, offering a natural framework for studying entangled interconnected systems across disciplines like neuroscience \cite{Fornito}, ecology \cite{Bascompte}, and epidemiology \cite{Pastor2001}, among others \cite{Newman2003}. Understanding the evolutionary dynamics of complex networks, as they adapt their connectivity patterns to achieve diverse goals, is crucial to understanding their long-term stability or other features influencing functional roles and performance \cite{Fortuna, Debian}. Notably, amidst the multitude of potential network structures, one organization ubiquitously arises in natural systems: the scale-free architecture \cite{Barabasi1999, NewmanBook,barabasi2016,caldarelli}.

Scale-free networks manifest a distribution of node connectivities $k$ that decays as a power-law $P(k) \propto k^ {-\gamma}$  for large values of $k$ \cite{Barabasi1999,Broido2019}. In statistical physics, power-law behavior is the hallmark of scale invariance and scaling behavior, implying no significant characteristic value of the analyzed quantity \cite{Newman-power,Sornette,MAMRMP}. For scale-free networks, there is no typical scale in the degree of connectivity apart from natural cut-offs. However, the network community has been intrigued by the possibility of discerning, through careful statistical analyses \cite{Clauset, Broido2019}, including finite-size effects \cite{Serafino2021}, whether empirical networks genuinely exhibit \emph{bona-fide} scale invariance or only appear to. From a theoretical standpoint, addressing this question calls for designing a renormalization group (RG) approach \cite{Wilson1979, Amit, JSTAT, KardarBook}. In this framework, a fixed point of the RG transformation denotes a state or system whose characteristics remain unchanged under appropriate scale transformations achieved through iterated coarse-graining \new{and rescaling}. RG fixed points are inherently linked to universal scaling laws governing the system, so systems at the same fixed point share identical scaling features and belong to the same universality class \cite{Amit, KardarBook}. Hence, a key implication of RG theory is the classification of numerous seemingly disparate physical systems and dynamical models into a relatively compact set of universality classes at criticality, whether in or out of thermal equilibrium \cite{HH,Odor2004}. Elucidating the relevant ingredients that give rise to a particular universality class offers valuable insights into the fundamental mechanisms underpinning complex systems and their key features.

Given the lack of a natural Euclidean embedding for complex networks, traditional length-scale transformations using translational invariance were deemed unfeasible until recently \cite{Radicchi, Rozenfeld2007}. In particular, small-world effects, characterized by short path lengths between nodes, further complicate block identification or affine transformations \new{in networks} \cite{Cohen2003, WS, Rozenfeld2007,Rozenfeld2010}. Nevertheless, various techniques have been proposed to tile networks \cite{Radicchi}. These include box-covering techniques \cite{Makse2005, Rozenfeld2010}, spectral partitioning \cite{Gfeller}, and hyperbolic geometry embeddings, which offer a promising novel approach for understanding particular complex network structures and dynamics \cite{Serrano2008, Serrano2018}. 
However, fully characterizing scale-invariant properties in real networks remains an open challenge.

A novel approach, the Laplacian Renormalization Group (LRG), generalizes RG to graphs, providing a comprehensive framework for coarse-graining heterogeneous systems in real and momentum space \cite{LRG}. This approach uses diffusive dynamics on networks to gradually remove the smallest fine-grained structural scales, by eliminating the contribution from large Laplacian eigenvalues. Specifically, the LRG allows for the redefinition of effective Laplacian and adjacency matrices \new{at coarse-grained scales}. Due to its general validity, it comes natural to explore structural scale invariance within the LRG framework, as evidenced by the observation of a time-independent rate of entropy loss \cite{LRG, Klemm2023}. This leads to the fundamental question of whether non-Euclidean architectures, like general networks, can be intrinsically scale-invariant. Two key questions arise. Can we identify benchmark classes of networks exhibiting genuine scale-invariant behavior and, thus, define an analog to universality classes?  Can we detect and quantify this kind of self-similarity in real-world networks? 
Here, we provide a clear definition and compilation of scale-invariant networks, emphasizing that scale-freeness, i.e., a power law degree distribution, does not imply structural self-similarity, and vice versa. Finally, we analyze real brain networks and highlight their scale-invariant features.

\paragraph{\textbf{Laplacian Renormalization Group (LRG).}}
Using a statistical mechanical approach, the LRG, allows for the detection of \new{relevant scales} in \new{weighted and undirected} networks \cite{LRG,PRX}. The LRG is based on the time-evolution operator \new{at time $\tau$}, $e^{-\tau\hat L}$ of the diffusion or heat equation, where  $\hat L=\hat D-\hat A$ is the Laplacian operator, $\hat A$ the adjacency matrix, and $\hat D$ the diagonal degree matrix \cite{InfoCore}. Using this and denoting the Laplacian eigenvalues as $\lambda_i$ with $i=1,...,N$ (all real and positive \cite{Chung1997,Ramanujan}), one can define the Laplacian density matrix \cite{Domenico2016},
\begin{equation}
 \hat \rho(\tau)=\frac{e^{-\tau \hat L}}{Z \equiv Tr (e^{-\tau \hat L})}=\frac{e^{-\tau \hat L}}{\sum_{i=1}^N e^{-\lambda_i\tau}}\,,
 \label{RhoMat}
\end{equation}
 and develop a "canonical" description of heterogeneous networks, fully analogous to that in statistical mechanics \cite{LRG, InfoCore}. Note that, $\hat L$ formally acts as a Hermitian Hamiltonian, $Z(\tau)$ is the partition function, and  $\tau$ serves as a control scale parameter analogous to the inverse temperature.  In this way, as the resolution scale $\tau$ is increased, the contribution of large eigenvalues to $\hat{\rho(\tau)}$ ---revealing fine structure--- is progressively removed, allowing for an effective network coarse-graining \footnote{This idea is also at the basis of different existing community detection algorithms  \cite{PRX,Arenas2006,Donetti-communities,Donetti-entangled}}. Thus, one can compute the network entropy as  $S(\tau)=-\mbox{Tr}[\hat \rho (\tau) \log \hat \rho (\tau)]$, 
 so that $S(\tau)\sim\tau \langle \lambda \rangle_\tau + \ln Z(\tau)$, where $\langle \lambda \rangle_\tau\equiv\langle \hat L\rangle_{\tau}=Tr[\hat\rho(\tau)\hat L] =
 \nicefrac{\stackrel[i=1]{N}{\sum}\lambda_i e^{-\tau\lambda_i}}{Z(\tau)}$. In particular, $S$ runs from $S=\ln N$ at $\tau=0$, the segregated regime to $S=0$ at $\tau \rightarrow \infty$, the fully integrated regime. \new{Moreover}, one can define the network entropic susceptibility or "heat capacity"  \cite{InfoCore}  as,
\begin{equation}
C(\tau)\equiv -\frac{dS}{d \log \tau}=- \tau^2 \frac{d\langle \lambda \rangle_\tau}{d\tau},
\label{SHeat}
\end{equation}
\new{describing} the rate of entropy loss (at which the complexity of the network shrinks upon coarse-graining) or the rate of information acquired about the network structure during diffusion dynamics at scale $\tau$. 
\new{In analogy with statistical physics}, peaks of $C$ (diverging in the infinite size limit) are associated with phase transitions. Thus, we can analyze $C$ at varying $\tau$ to investigate the network multi-scale organization, detecting scales where entropy changes more significantly due to structural transitions. This description allowed, for example, the detection of the network information core and its associated structural and diffusive transitions \cite{InfoCore}, and led to the natural extension of RG to heterogeneous networks \cite{LRG}. Here, it allows us to define informationally {\em scale-invariant networks} as graphs whose entropy-loss rate $C(\tau)$ takes a constant value $C_0>0$ across scales or, at least, within a sufficiently broad diffusion-time interval, thus being (exactly or approximately) scale invariant.

We demonstrate that this definition of informational scale-invariance holds if the Laplacian spectral density follows $P(\lambda) \sim \lambda^{\gamma}$, considering it as a continuum distribution in the infinite network-size limit. Using Eq.\eqref{SHeat} and setting $C(\tau)=C_0$ as a constant, one gets $\langle \lambda \rangle_\tau=C_0/\tau$. The only solution to this equation is a Laplacian spectral density $P(\lambda) \sim \lambda^{\gamma}$, \new{which allows  us to express} $C_0$ as a function of the exponent $\gamma$, specifically $C_0=\gamma+1$. This implies that macroscopic properties  ---stemming from \new{small} Laplacian eigenvalues--- of a network with a power-law spectrum remain invariant under LRG transformations. Thus, \new{the concept of} network scale-invariance shifts from a power-law degree distribution to a power-law in the spectral density. Note that the exponent of the Laplacian spectral density, \(\gamma\), can be related to the \emph{spectral dimension}, \(d_s\), which is a global property of the graph, related, e.g., to the infrared singularity of the Gaussian process \cite{Burioni1996} and that has been shown to provide a robust generalization of the standard concept of dimension for networks \cite{Cassi1992,Burioni1996}. As $d_s/2 = \gamma + 1 $ \cite{Burioni1996}, one concludes that \(C_0\) is constant and equals half of the graph's spectral dimension: \(C_0 = d_s/2\). Thus,  measuring the plateau value of the heat capacity of a scale-invariant network effectively determines its dimensionality.

For finite-sized networks, scale invariance  may only be approximate and confined to a finite-scale interval. This means that \(C(\tau)\) cannot remain constant across all scales: it decays from its plateau value \(C_0\) to \(0\) for sufficiently large values of \(\tau\). To estimate this cut-off, we have to consider that the smallest non-zero eigenvalue $\lambda_2$ of $\hat L$, also called "spectral gap" or "Fiedler eigenvalue" $\lambda_F$, \cite{Fiedler1989, Chung2003}, is the one providing the slowest decaying contribution in time to Eq.\eqref{RhoMat}, therefore determining the asymptotic time decay of \(C(\tau)\). Starting from Eq.\eqref{SHeat} and imposing that the Laplacian spectrum $P(\lambda)$ integrates to $N$, one obtains $\lambda_F\sim N^{\nicefrac{-2}{d_s}}$ for large $N$ \cite{L-spectrum,DoroBook}. From this, it follows that \new{for large times $\tau_F(\alpha)=\alpha/\lambda_F$, where $\alpha\gg 1$ is a free parameter}
(see Supplemental Material, SM \cite{SM})
\begin{equation}
{\small C\left(\tau_F(\alpha)\right) \simeq \frac{\Gamma\left(\frac{d_s}{2}\right)\Gamma\left(\frac{d_s}{2}+2,\alpha\right) - \Gamma^2\left(\frac{d_s}{2}+1,\alpha\right)}{\Gamma^2\left(\frac{d_s}{2}\right)},}
\label{Scaling}
\end{equation}
where $\Gamma(x,\alpha)=\int_{\alpha}^\infty du\,u^{x-1}e^{-u}$ is the lower incomplete Euler Gamma function. \new{This allows a complementary way to estimate $d_S$ from finite-size scaling analyses of $C(\tau_F)$ in finite networks, by studying the scaling of the $\tau_F$ values for which $C(\tau_F)$ falls below some given threshold value (e.g., $C(\tau_F)=\frac{1}{2}$) as a function of $N$.}

{\textbf{Laplacian Random-Walk RG.}} The LRG method relies on a ``heat-like" diffusion process defined by $\hat L$. However, we wonder whether the results obtained for scale-invariance and graph dimensionality remain valid when using the Laplacian random-walk operator, $\hat L_{RW}=\hat D^{-1}\hat L$ instead of $\hat L$. $\hat L_{RW}$ describes the time-discrete dynamics of a RW moving from a node to a neighbor with uniform probability and is related to the transition matrix for RW dynamics on a graph \cite{Burioni2005, Chung1997}. Although $\hat L_{RW}$ is not symmetric, it is similar to the symmetric operator $\hat L_{sym}=D^{1/2}\hat L_{RW}\hat D^{-1/2}$ \cite{Chung1997} with all its eigenvalues real and satisfying $0\le \mu_i\le 2$. We can, thus, reformulate the LRG for the RW dynamics by substituting $\hat L$ with $\hat L_{sym}$, renaming the related heat capacities as $C_{L}$ and $C_{RW}$, respectively. In fact, $\hat L_{RW}$ provided the original definition of the RW spectral dimension, $d_s^{RW}$, related to the first-return time distribution of the random-walker $P_0(t)$ \cite{Burioni1996, Cassi1999} by the scaling $P_0(t)\propto t^{-d_s^{RW}/2}$. This generalizes the behavior on $d-$dimensional lattices where $P_0(t)\propto t^{-d/2}$ \cite{alexander1982density}. While $\hat L$ or $\hat L_{RW}$ strictly coincide in regular networks like lattices, both operators unravel heterogeneous network structures in different ways \cite{PRX} and, to our knowledge, there is no proof that $d_s=d_s^{RW}$ for generic heterogeneous networks.

{\textbf{LRG non-trivial fixed points.}}
We now tackle the following lingering questions: What are the current non-trivial fixed points of the LRG, i.e., families or "universality classes" of scale-invariant networks? Are scale-invariant networks necessarily scale-free? Do results depend on the choice of the operator?  

Let us remark that regular lattices, with a degree distribution $P(\kappa)=\delta(\kappa-\kappa_0)$, are the simplest case of scale-invariant structures lacking scale-free properties (see SM \cite{SM}). However, our main focus is the categorization of \emph{heterogeneous} and \emph{stochastic} scale-invariant networks.
\begin{figure}[hbtp]
    \centering
\includegraphics[width=\columnwidth]{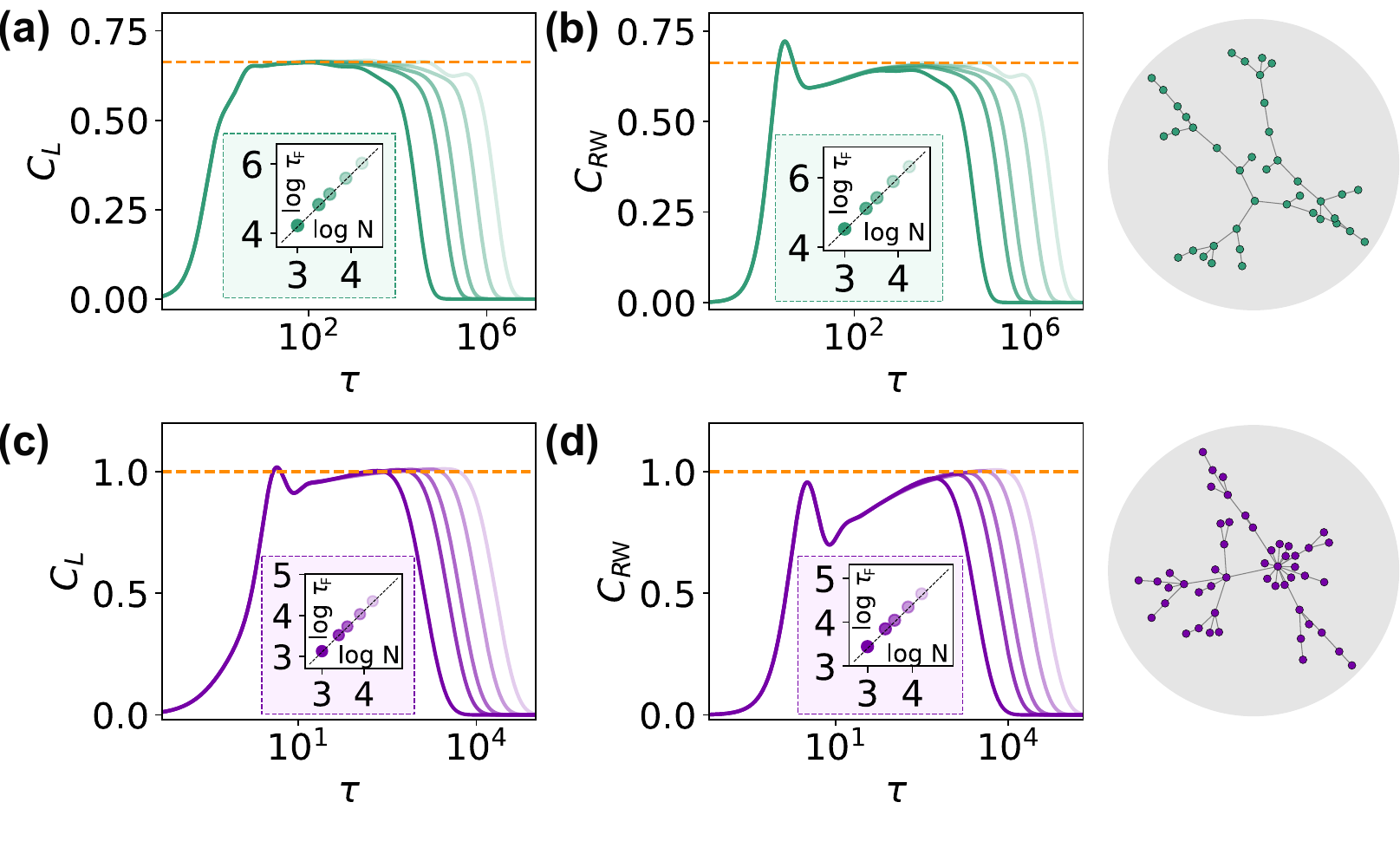}
\caption{\textbf{Trees.} \new{Heat capacity as a function of the resolution scale $\tau$}: \textbf{(a)-(b)} RTs and \textbf{(c)-(d)} BA networks with $m=1$. We use $\hat L$ in \textbf{(a)}  and \textbf{(c)} and $\hat L_{RW}$ in \textbf{(b)} and \textbf{(d)}. Insets show the scaling of $\tau_F$ as a function of network size $N$. Orange dashed lines and solid black lines represent the \new{scaling for the }theoretical expectation, $d_s^{RT}=\nicefrac{4}{3}$ and $d_s^{BA}=2$. All curves are averages over $10^3$ independent realizations. Different colors represent different sizes $N = \{1, 2.5, 4, 8, 16\} \times 10^3$.}
\label{Fig.BT}
\end{figure}

The \textbf{first} category of self-similar networks comprises trees, i.e., connected loopless networks. It is known that the spectral dimension of random trees, with minimal branching ratio $b_{min}=1$, depends upon the first two moments of the degree distribution $ P(\kappa) $ \cite{donetti2004statistical, Burda2001}. Specifically, $d_s = 4/3 $ if $ \langle \kappa ^2 \rangle $ is finite, while the problem remains open when $ \langle \kappa ^2 \rangle $ diverges \cite{donetti2004statistical,destri2002spectral, dorogovtsev2003}. 
Here, we examine specific cases within these two classes, using the lenses of LRG: ordinary random trees (RT) and scale-free Barabási-Albert (BA) networks where new nodes attach preferentially to existing ones, forming $m=1$ edges \cite{Barabasi1999}.  As reported in Fig.\ref{Fig.BT}(a), $C_L$ shows a plateau corresponding to the theoretically known spectral dimension for RTs, $ d_s^{RT}=\nicefrac{4}{3} $ \cite{destri2002spectral} and obeys the finite-size scaling condition, Eq.\eqref{Scaling} \new{for various network sizes}. We also observe that $ \hat L_{RW} $ does not alter this value (see Fig.\ref{Fig.BT}(b)), despite specific differences in the $C$ shape, as the local peak at short times in Fig.\ref{Fig.BT}(b).

Instead, scale-free BA networks with $m=1$ are trees with diverging $\langle \kappa ^2 \rangle$. As shown in Fig.\ref{Fig.BT}(c), they reach a constant heat capacity $C_L$ corresponding to a spectral dimension $d_s=2$. $C_{RW}$ grows in the intermediate regime but becomes flat for asymptotic times and large network sizes (see Fig.\ref{Fig.BT}), confirming scale-invariance with $d_s^{RW}=d_s=2$. This same spectral dimension is also obtained for Bethe lattices with coordination number $z\ge 3$ (see \cite{Erzan2011} and SM \cite{SM}). Furthermore, BA networks with $m>1$ show no sign of scale invariance (see SM \cite{SM}), proving that the archetypes of scale-free networks are not scale-invariant. This prevents preferential attachment from generating self-similar networks with $d_s>2$. 

\begin{figure}[t]
    \centering
\includegraphics[width=\columnwidth]{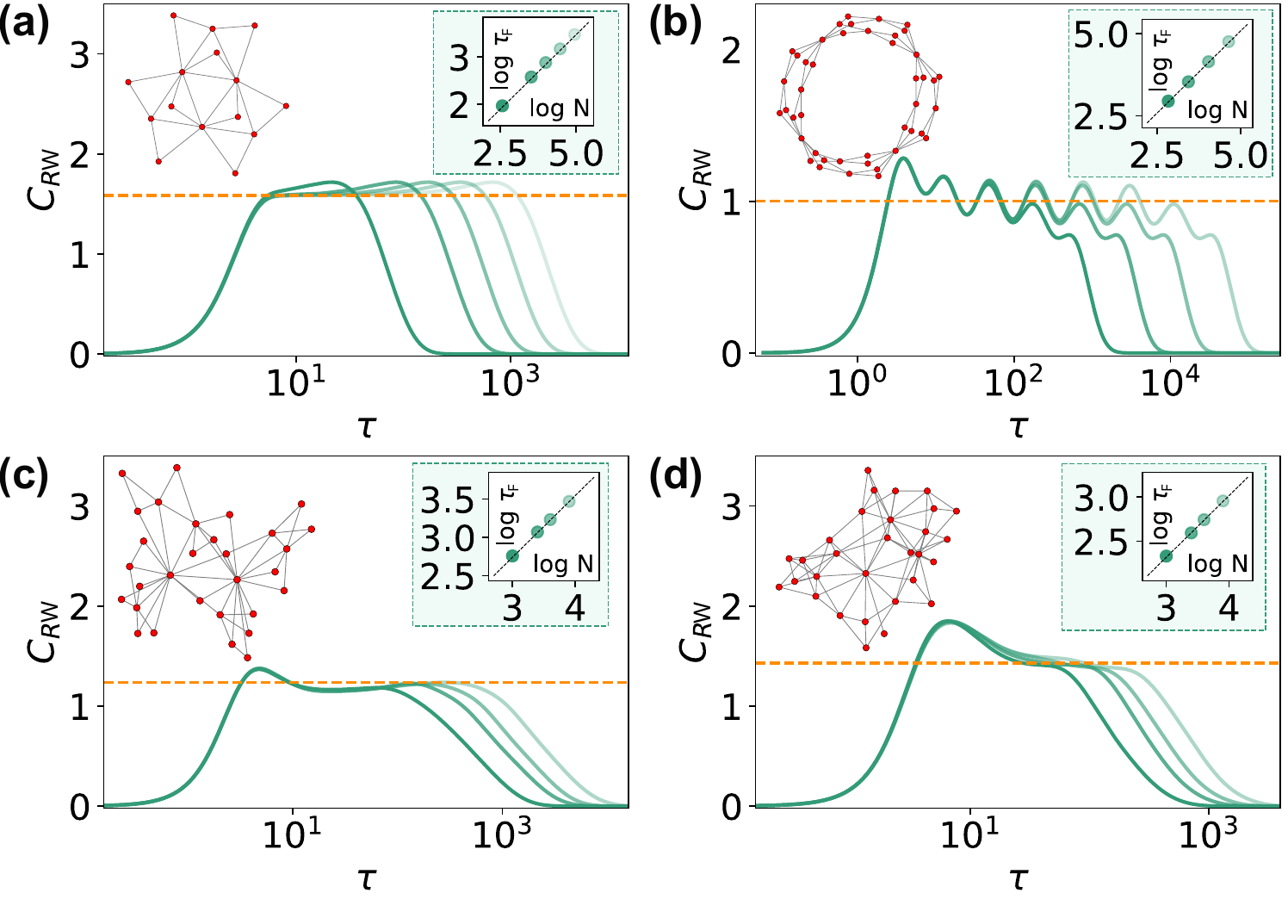}
\caption{\textbf{Clustered networks.} \new{$C_{RW}$ as a function of the resolution scale $\tau$ for}: \textbf{(a)} $(1,2)$ flowers with  $s=6, 8, 9, 10$ and $11$, hierarchical levels, respectively. \textbf{(b)} $(2,2)$ flowers with $s=5, 6, 7$, and $8$ levels resp. For $(u,v)$ flowers, the number of nodes is $N_s=\left(\frac{w}{w-1}+\frac{w-2}{w-1}w^s\right)$, with $w=u+v$. \textbf{(c)} KH networks with $m=2$ and \textbf{(d)} KH with $m=3$ of sizes $N = \{1, 2.5, 4, 8\} \times 10^3$. Insets show the scaling of $\tau_F$ as a function of network size $N$. Orange dashed lines and solid black lines represent the theoretical expectation. All curves have been averaged over $10^3$ independent realizations.}
\label{Fig.Flowers}
\end{figure}

\textbf{Second}, we consider networks that, contrary to trees, have a non-vanishing clustering coefficient. In particular, Dorogovtsev-Goltsev-Mendes \cite{dorogovtsev2002} networks and their generalization: \emph{(u,v)-flowers} \cite{dorogovtsev2002, Rozenfeld2007, Rozenfeld2008}. These deterministic fractal-like structures grow by iteratively replacing the link between two nodes with two paths of lengths $u$ and $v$, respectively (see SM \cite{SM}). Diameter-based analyses, such as calculating the Hausdorff dimension of these networks lead to an apparent paradox: all $(1,v)$ flowers are infinite-dimensional and exhibit anomalous scaling functions \cite{Rozenfeld2007,dorogovtsev2002,peng2017scaling}. This paradox stems from the lack of network embedding in an Euclidean space. Conversely, as shown in Figs.\ref{Fig.Flowers}(a/b) and in the SM \cite{SM}, all flowers exhibit well-defined spectral dimensions \cite{Rozenfeld2007}, $d_s=2\log(\emph{u}+\emph{v})/\log(\emph{u}\emph{v})$ changing from $d_s\approx3.17$ to $d_s=1$.
Fig.\ref{Fig.Flowers}(b) illustrates that $(u,v)$ flowers with $u>1$ exhibit characteristic oscillations emerging from the discrete nature of the recursive network growth process (as log-periodic oscillations \new{stemming from discrete-scale invariance} in standard fractal analyses \cite{Sornette,PietroneroBook}). This demonstrates that the \new{RG method is also able to detect the presence of} discrete scale invariance on networks.

\textbf{Third,}
to explore self-similarity in stochastic scale-free graphs with small-world properties \cite{Rozenfeld2007}, we consider the Kim-Holme (KH) network model \cite{KimHolme}. This modifies BA networks by introducing a probability $p$ for each newly added node to form a triangle with existing nodes (see SM \cite{SM}). The resulting networks show a power-law degree distribution and a high clustering coefficient. LRG analysis of KH networks reveals scale-invariance only for $p=1$. As shown in Figs.\ref{Fig.Flowers}(c) and (d), $C_{RW}$ (and $C_L$ \cite{SM}) shows a plateau with spectral dimension growing with $m$ from $d_s=2.57(1)$ to $d_s=3.65(1)$ (see SM \cite{SM}). 
 
\begin{figure}[t]
    \centering
\includegraphics[width=\columnwidth]{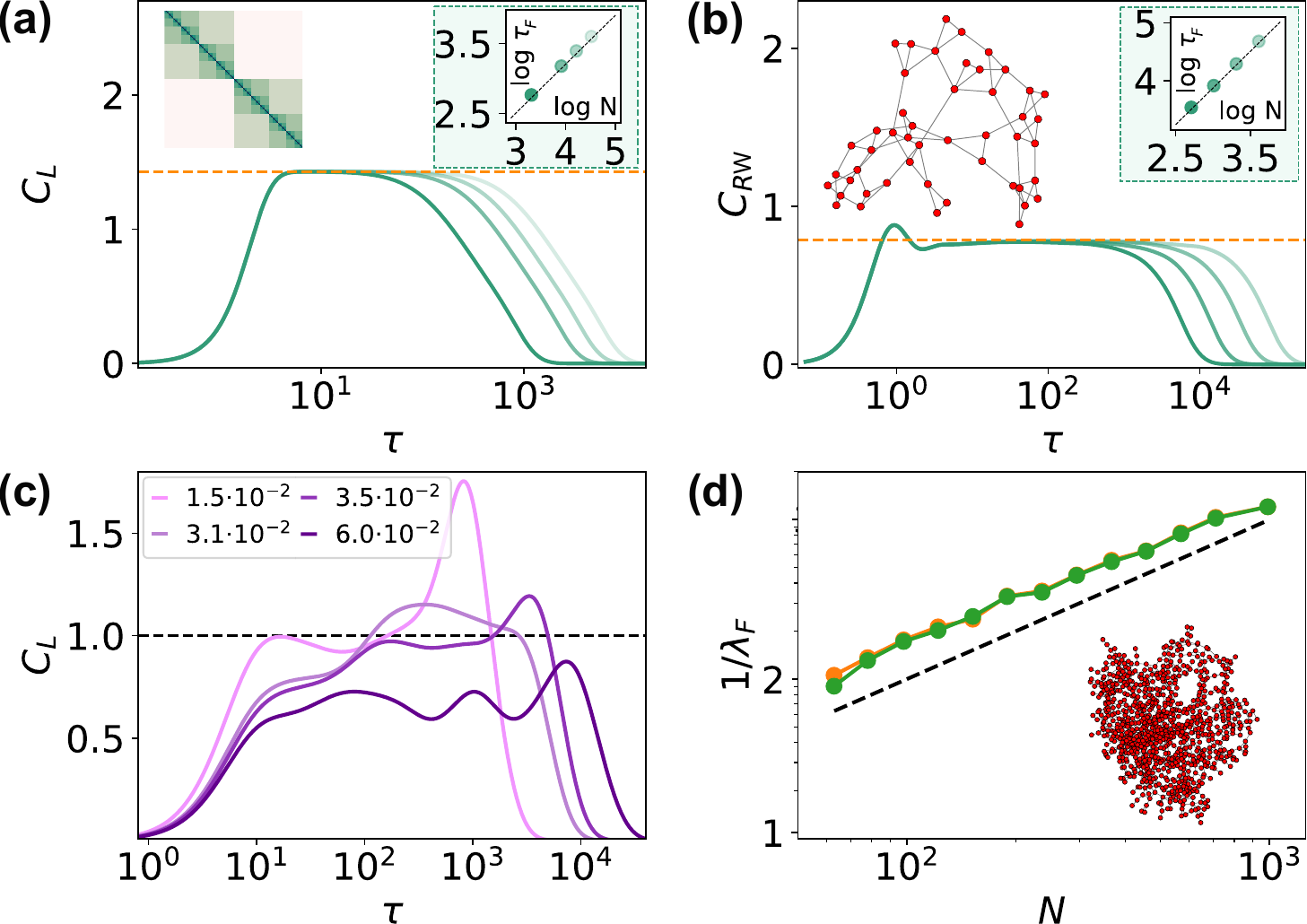}
\caption{\textbf{Hierarchical networks.} Heat capacity  as a function of the resolution scale $\tau$ for: \textbf{(a)} Dyson graphs with 
$\sigma=0.85$ and $s=11, 13, 14, 15$ hierarchical levels resp.(network size is $N=2^s$ and the upper left inset shows the weighted adjacency matrix); \textbf{(b)} HMNs with $m_0=3$ and $\alpha=2$ (see SM \cite{SM}) for  $s=9, 10, 11, 12$ hierarchical levels resp. ($N=2^sm_0$). Right insets \textbf{(a)} and \textbf{(b)} show the scaling of $\tau_F$ as a function of $N$. Orange dashed and solid black lines, respectively, show the theoretical expectation. \textbf{(c)} HC network for different thresholds (see legend). \textbf{(d)} \new{Scaling of $\lambda_F$ for coarse-grained versions of the HC with $T=3\cdot10^{-2}$ and $\tau-$values of $10^{-1}$, $10^0$, and $10^1$, which are superimposed. Black dashed line represents the scaling using $d_s=1.9$.}}
    \label{Dyson}
\end{figure}

\textbf{Fourth,} we investigate networks with a built-in hierarchical structure. In particular, we analyze the Dyson graph \cite{dyson1969existence,agliari2017}: a fully-connected deterministic graph with hierarchically organized link weights, resulting in a tunable spectral dimension $d_s=2/(2\sigma-1)$, being $\sigma$ a scaling parameter controlling the weight strength at every scale (see SM \cite{SM}). Fig.\ref{Dyson}(a) shows the constant plateau of $C_L$ for a Dyson graph with $\sigma=0.85$, consistent with the theoretical value. 
We have also examined hierarchical modular networks (HMNs) that were proposed using inspiration from brain networks \cite{moretti2013griffiths}. In HMNs, nodes grouped in basal fully-connected modules are recursively coupled with nodes in other moduli, establishing inter-modular links randomly in a hierarchical fractal-like manner  (see SM \cite{SM}). Fig.\ref{Dyson}(b) shows their heat capacity: HMNs are scale-invariant for any set of parameters, with spectral dimension in the range $d_s\in(1.25,2)$ (see SM \cite{SM}).

\textbf{Finally}, we have analyzed the Human Connectome (HC) structural brain network \cite{Fornito} (see SM \cite{SM}). \new{To account for fluctuations in single realizations of empirical networks, we apply the LRG \cite{LRG} to analyze reduced HC versions, performing finite-size scaling (see SM for the detailed procedure \cite{SM}). Due to the high density of weak links, the HC must be sparsified by removing links below a threshold (T) to reveal non-trivial scale-invariant features, but avoiding the trivial RT expected at the percolation critical point (at $T \approx 0.06$ in this case). Rusults of the G  analysis of $C_L$ for thresholded HC networks is shown in Fig.\ref{Dyson}(c), while Fig.\ref{Dyson}(d) illustrates the robust} finite-size scaling of $\lambda_F$ for reduced versions of a HC \new{at different $\tau$-values}. Both analyses support an underlying scale-invariant topology  with spectral dimension $d_s \sim 1.9(1)$  \new{(in agreement with \cite{Serrano2020})}.


\paragraph{\textbf{Outlook.}}

The observation of scale-invariance across length and time scales led 
to the development of RG ideas \cite{KardarBook,Amit} whose initial implementation took advantage of the geometrical homogeneity of interactions and translational invariance of the \new{embedding space} to perform scaling of statistical physical systems, computing critical exponents \cite{Wilson1972}, and leading to classify apparently diverse phase transitions into a few universality classes  \cite{HH,Odor2004}.

Extending universality to networks with heterogeneous and non-local connectivity properties has long challenged physicists. Despite remarkable advances have been recently made \cite{Makse2005,Serrano2008,Rozenfeld2010,Serrano2018,Serrano2020}, fully understanding scale-invariance in generic networks remains unresolved. Initially, network scale-invariance was associated with power-law degree distributions \cite{Serafino2021}. However, as we explicitly show, {\em scale-freeness} of node degrees is neither necessary nor sufficient for true self-similarity \cite{Broido2019}.

Instead, the LRG \cite{LRG,InfoCore} provides a framework \new{capable of} characterizing and classifying \emph{non-trivial} structural fixed points, describing their scaling properties and offering a classification in terms of universality classes.
 A constant heat-capacity indicates the presence of scale-invariance or self-similarity in graphs, characterized by a constant entropy-loss rate during network coarse-graining.  We have linked the constant entropy-loss rate to the network spectral dimension, $d_s$ ---computed using either $\hat L$ or $\hat L_{RW}$--- confirming its role as a natural generalization of the Euclidean dimension \cite{Cassi1992, Cassi1999, Burioni1996}. 

Our LRG analysis identifies regular lattices, trees, clustered networks, and hierarchical networks as fundamental classes of scale-invariant networks. Beyond regular lattices, we highlight (u-v)-flowers and KH networks as unique structures combining local clustering with hubs at all scales to generate self-similar topologies. Trees are bound to $d_s\leq 2$, and simply adding hubs appears insufficient to achieve \new{large dimension values}. Generally, as the dimension increases, it becomes more challenging to generate self-similar heterogeneous networks \cite{Rozenfeld2007}.  Our findings are promising for studying biological and socio-technological networks; notably, hierarchical networks, as the HC \cite{Sporns2004,Hagmann2008} which exhibit robust scale invariance with $d_s\approx1.9(1)$, corroborating previous results \cite{Serrano2020}. We provide a solid ground for future dynamical RG theoretical calculations on these structures and for identifying new signatures of scale invariance in real networks.
\vspace{-0.5cm}
\begin{acknowledgments}
\section*{Acknowledgments}
\vspace{-0.2cm}
We thank T. Gili, D. Cassi, and R. Burioni for very useful discussions and comments. P.V. and M.A.M. acknowledge financial support from the Spanish "Ministerio de Ciencia,  Innovaci\'on y Universidades" and the "Agencia Estatal de Investigaci\'on (AEI)" under Project Ref. PID2020-113681GB-I00 funded by  MICIN/AEI/10.13039/501100011033 and  
Grant No. PID2023-149174NB-I00 financed by MICIU/AEI/10.13039/501100011033 and EDRF/EU funds.

\end{acknowledgments}
\vspace{-0.5cm}

\def\url#1{}
%
\end{document}